\documentclass[reprint,amsmath,amssymb,groupedaddress,aps]{revtex4-1}

\usepackage{graphicx}	
\usepackage{subcaption} 
\usepackage{wasysym}
\usepackage{dcolumn}	
\usepackage{bm}			
\usepackage{color,soul}
\usepackage{xcolor}		
\usepackage{hyperref}

\begin{document}

\newcommand{\perrthz}[0]{\,Hz$^{-1/2}$}
\newcommand{\oneOverF}[0]{1\!/\!$f$}
\newcommand{\idunno}[0]{\fcolorbox{red}{blue}{\color{white} 666}}
\newcommand{\lolwut}[1]{\fcolorbox{red}{blue}{\color{white} #1}}
\newcommand{\rubidium}[0]{$^{87}$Rb}
\newcommand{\uW}[0]{~$\mu$W}
\newcommand{\uS}[0]{~$\mu$s}
\newcommand{\mW}[0]{~mW}
\newcommand{\mS}[0]{~ms}
\newcommand{\fthz}[0]{~fT\perrthz{}}
\newcommand{\pthz}[0]{~pT\perrthz{}}
\newcommand{\brf}[0]{$\boldvec{B}_{\text{rf}}$}
\newcommand{\brfm}[0]{\boldvec{B}_{\text{rf}}}
\newcommand{\bmain}[0]{$\boldvec{B}_0$}
\newcommand{\bcomb}[0]{$\boldvec{B}_{\pi}$}
\newcommand{\temp}[1]{$#1^{\circ{}}\!\text{C}$}
\newcommand{\F}[1]{$F=#1$}
\newcommand{\FF}[2]{$F=#1 \rightarrow F'=#2$}
\newcommand{\xz}[1]{\fcolorbox{red}{blue}{\color{white}#1}}
\newcommand{\xzcite}[0]{\xz{[?]}}
\newcommand{\lolrly}[0]{\xz{RLY?}}
\newcommand{\axis}[1]{\textbf{\^{#1}}}
\newcommand{\parVec}[2]{#1\,$\parallel$\,#2}
\newcommand{\perpVec}[2]{#1\,$\perp$\,#2}
\newcommand{\kprobe}[0]{$\boldvec{k}_{\mathrm{pr}}$}
\newcommand{\kpump}[0]{$\boldvec{k}_{\mathrm{pmp}}$}
\newcommand{\kls}[0]{$\boldvec{k}_{\mathrm{LS}}$}
\newcommand{\hlc}[2][yellow]{ {\sethlcolor{#1} \hl{#2}} }
\newcommand{\pipulse}[0]{$\pi$-pulse}
\newcommand{\fig}[1]{Fig.~\ref{#1}}
\newcommand{\fmod}[0]{$f_{\mathrm{mod}}$}
\newcommand{\boldvec}[1]{\bm{#1}}
\newcommand{\boldunitvec}[1]{\bm{\hat{#1}}}

\newcommand{\pipulseProbeNoiseXlowNum}[0]{0.7}
\newcommand{\pipulseProbeNoiseYlowNum}[0]{1.5}
\newcommand{\pipulseProbeNoiseXhighNum}[0]{2}
\newcommand{\pipulseProbeNoiseYhighNum}[0]{8}
\newcommand{\dcSERFProbeNoiseHighNum}[0]{27}
\newcommand{\dcSERFProbeNoiseLowNum}[0]{0.3}
\newcommand{\ZmodProbeNoiseXlowNum}[0]{0.5}
\newcommand{\ZmodProbeNoiseYlowNum}[0]{0.5}
\newcommand{\ZmodProbeNoiseXhighNum}[0]{1}
\newcommand{\ZmodProbeNoiseYhighNum}[0]{59}
\newcommand{\pipulseProbeNoiseXlow}[0]{\pipulseProbeNoiseXlowNum{}\fthz{}}
\newcommand{\pipulseProbeNoiseYlow}[0]{\pipulseProbeNoiseYlowNum{}\fthz{}}
\newcommand{\pipulseProbeNoiseXhigh}[0]{\pipulseProbeNoiseXhighNum{}\fthz{}}
\newcommand{\pipulseProbeNoiseYhigh}[0]{\pipulseProbeNoiseYhighNum{}\fthz{}}
\newcommand{\dcSERFProbeNoiseHigh}[0]{\dcSERFProbeNoiseHighNum{}\fthz{}}
\newcommand{\dcSERFProbeNoiseLow}[0]{\dcSERFProbeNoiseLowNum{}\fthz{}}
\newcommand{\ZmodProbeNoiseXlow}[0]{\ZmodProbeNoiseXlowNum{}\fthz{}}
\newcommand{\ZmodProbeNoiseYlow}[0]{\ZmodProbeNoiseYlowNum{}\fthz{}}
\newcommand{\ZmodProbeNoiseXhigh}[0]{\ZmodProbeNoiseXhighNum{}\fthz{}}
\newcommand{\ZmodProbeNoiseYhigh}[0]{\ZmodProbeNoiseYhighNum{}\fthz{}}

\newcommand{\noiseHighFreq}[0]{0.01~Hz}
\newcommand{\noiseLowFreq}[0]{10~Hz}
\newcommand{\samplingrateFPGA}[0]{500~ksps}
\newcommand{\samplingrateNIDAQ}[0]{1~Msps}
\newcommand{\pumpPower}[0]{12\mW{}}
\newcommand{\probePowerDCZ}[0]{400\uW{}}
\newcommand{\probePowerPi}[0]{800\uW{}}
\newcommand{\noiseRecord}[0]{0.16\fthz{}}
\newcommand{\gratioPiPulse}[0]{5.56~Hz/nT} 
\newcommand{\slowingDownFactorPiPulse}[0]{5.7}
\newcommand{\polarizationPiPulse}[0]{0.29}
\newcommand{\bmainValuePiPulse}[0]{44~nT}

\newcommand{\uwmadison}[0]{\affiliation{Department of Physics, University of Wisconsin-Madison, 1150 University Avenue, Madison, Wisconsin 53706, USA}}

%

\title{Dual-axis pi-pulse spin-exchange relaxation-free magnetometer}

\author{Elena Zhivun}
\email{zhivun@gmail.com}
\uwmadison{}

\author{Michael Bulatowicz}
\email{bulatowicz@wisc.edu}
\uwmadison{}

\author{Alexander Hryciuk}
\email{hryciuk@wisc.edu}
\uwmadison{}

\author{Thad Walker}
\email{thad.walker@wisc.edu}
\homepage[]{www-atoms.physics.wisc.edu}
\uwmadison{}




\date{\today}

\begin{abstract}
We present a new spin-exchange relaxation-free vector magnetometer with suppressed \oneOverF{} probe noise, achieved by applying a small DC bias field and a comb of magnetic DC $\pi$ pulses along the pump direction.
This results in a synchronous orthogonal AC response for each of its two sensitive axes.  
The new magnetometer is particularly well-suited to applications such as biomagnetism in which the signal to be measured carries a dominant component of its power at low frequencies.
The magnetometer reaches a technical noise floor of 8.4\fthz{} (\axis{x}) and 11\fthz{} (\axis{y}) at 0.01~Hz. 
A single-axis DC SERF sharing the same experimental apparatus attains 61\fthz{} at the same frequency. 
A noise minimum of 1.1\fthz{} (\axis{x}) and 2.0\fthz{} (\axis{y}) is reached by the new magnetometer at 10~Hz, compared to 0.7\fthz{} at 25~Hz for a DC SERF.
\end{abstract}


\maketitle

\section{Introduction}

Precision measurement of weak magnetic fields can yield important information not obtainable by other methods.
Residual magnetization of geological samples reveals the Earth's magnetic field history, formation and movement of the continents, and provides means to verify geophysical theories~\cite{earth_mag_foundation,env_magnetism}.
The magnetic fields generated by electrical signals in the human body are used in both research and clinical diagnosis. 
Fetal magnetocardiography (fMCG), for example, is an important tool for diagnosing arrhythmia in a developing fetus \emph{in utero}~\cite{Strasburger2008}. 
Electric fetal heart signals are attenuated and distorted by the surrounding tissue and \emph{vernix caseosa} (a waxy substance covering the fetus), making electrocardiography (ECG) challenging~\cite{Wakai2000_vernix_caseosa}.
Similarly, magnetoencephalography (MEG), used to detect and localize brain responses to external stimuli or diagnose and localize pathological activity~\cite{book_Plonsey,meg_epilepsy}, offers better source localization and complementary information to electroencephalography (EEG).

While superconducting quantum interference devices (SQUIDs) have been an established state-of-the-art tool for these applications, optical atomic magnetometers are becoming a viable alternative.
These magnetometers are compact, reach similar magnetic sensitivity levels ($\sim$1\fthz{})~\cite{squid_handbook, DimaOpticalMagnetometry} and do not require liquid helium or a large magnetically shielded room, substantially reducing the cost of operation and potentially making high-sensitivity magnetometers more accessible in the future. 
Optical atomic magnetometers have been employed in high sensitivity measurements of remnant rock magnetization as a function of temperature~\cite{recordSERF}, brain auditory response~\cite{meg_romalis,meg_kitching,meg_cs}, multi-channel MEG~\cite{opm_meg_20ch,meg_Johnson2013}, and fMCG signal measurements competitive with SQUIDs~\cite{Wyllie2012,knappe_fmcg_array}.

Spin-exchange relaxation-free (SERF) magnetometers~\cite{firstSERF} are a subtype of optical atomic magnetometers which exhibit exceptional sensitivity (record of \noiseRecord{})~\cite{recordSERF}, making them particularly attractive for fMCG.
While traditional DC SERF magnetometers suffer from \oneOverF{} noise that can dominate the fMCG signal, this can be mitigated by adding an external magnetic field modulation, which facilitates signal detection at a higher frequency~\cite{elliptic_serf}.
Typically SERF magnetometers measure a single field vector component orthogonal to the optical pumping (\axis{z}) and probing (\axis{x}) axes [\fig{fig:setup}(a)].
Other field components can be measured by adiabatically modulating the magnetic field at the cost of drastically reducing the bandwidth~\cite{three_axis_serf_Seltzer2004}.
An alternate approach (``Z-mode'') is to modulate the field along \axis{z} at the frequency \fmod{} outside of the magnetometer's bandwidth~\cite{Zmod}.
Demodulating the signal at \fmod{} provides an independent measurement of the field along \axis{x}, while the \axis{y} component is detected either at DC with gain comparable to the \axis{x} component, or at 2\fmod{} with a substantially reduced gain.
In our system, better fMCG measurements~\cite{DeLandThesis} are achieved through both diffusive suppression of the AC Stark shifts~\cite{Sulai2013} and detection of the \axis{y} component at DC. 
However, this renders the \axis{y} field measurement prone to \oneOverF{} technical noise, degrading the \axis{y} sensitivity at low frequencies. 
While it is possible to circumvent this issue by introducing another probe beam along the \axis{y} axis~\cite{dual_probe_Zmod_Li2017,SeltzerThesis}, this requires three orthogonal optical axes and increases the complexity of each individual sensor.

We present here a new method for measuring both \axis{x} and \axis{y} field components using synchronous detection with a single probe, while retaining high sensitivity and spin-exchange relaxation suppression.
This is achieved by applying a superposition of a DC offset field \bmain{} and a comb of \pipulse{}s \bcomb{} to the sensor in the \axis{z} direction.
The signals produced by both \axis{x} and \axis{y} magnetic field components are periodic at  the \pipulse{} frequency $f_{\pi}$ and orthogonal to each other. 
The signal demodulation for each axis is performed in real time via multiplying the probe polarization rotation signal by an appropriately phased square-wave at the \pipulse{} frequency, followed by low-pass filtering.
This approach suppresses \oneOverF{} technical noise along the  \axis{y} axis in the \pipulse{} magnetometer, as compared to the DC SERF and the Z-mode.
The technical noise limit reached by the \pipulse{} magnetometer is comparable to the Z-mode magnetometer, while the DC SERF magnetometer attains a lower noise limit due to the higher Faraday rotation gain.

\section{Theory}
Consider a spin ensemble in the magnetic field $\boldvec{B} = \boldvec{B}_0 + \boldvec{B}_{\perp} + \boldvec{B}_{\pi}$, where $\boldvec{B}_{\perp} = B_x \boldunitvec{x} + B_y \boldunitvec{y}$ is the field to be measured, and $\boldvec{B}_0$ is the offset field in the \axis{z} direction, which is parallel to the pump wave vector \parVec{\kpump{}}{\bmain{}}. 
A comb of short \pipulse{}s \parVec{\bcomb{}}{\bmain{}} has repetition rate $f_{\pi}$. 
Here \pipulse{} is defined as a magnetic field pulse causing the atomic spin vector $\boldvec{S}$ to undergo Larmor precession by the angle $\pi$ around \axis{z}.
Let $\Omega_+ = \gamma B_{+}$ and $\Omega_0 = \gamma B_{0}$ be the corresponding precession rates in the constant magnetic field, where $\gamma$ is the gyromagnetic ratio, and $B_+ = B_x + i B_y$.
The Bloch equation in the spherical basis for the $S_+ = S_x + i S_y$ component is
\begin{equation}
\frac{dS_+}{dt} = \left[-\Gamma +i \left(\pi  f_{\pi }+\frac{d\phi _{\pi }}{dt} + \Omega_0\right)\right]S_+-i \Omega_+ S_z,
\label{eqn:bloch}
\end{equation}
where $\Gamma$ is the spin relaxation rate, $\gamma B_{\pi} = d\phi_{\pi}/dt$, and $\phi_{\pi}$ is defined as $\phi_{\pi} \equiv -\pi \left[ f_{\pi } t \pmod 1 \right]$.
After the transformation $S_+ = A_+ e^{i \phi _{\pi }}$, Eq.~\ref{eqn:bloch} is simplified: 
\begin{equation}
\frac{dA_+}{dt}=
	\left[
		-\Gamma +i \left(\pi  f_{\pi } + \Omega_0 \right)
    \right] A_+ - i \Omega_+ S_z e^{-i \phi_{\pi}}.
\label{eqn:bloch2}
\end{equation}
Since $e^{-i \phi_{\pi}}$ is a periodic function, the resonance condition can be found by substituting $A_+ = \sum _p A_{(+,p)} e^{i p t \omega_{\pi }}$, 
$e^{-i \phi_{\pi }}= \sum _p j_p e^{i p t \omega _{\pi }}$ in steady state:
\begin{equation}
A_{(+,p)}=-\frac{ i \Omega _+ j_p S_z}{\Gamma -i \left(\pi  f_{\pi } - p \omega _{\pi }+\Omega _0\right)}, 
\label{eqn:resonance}
\end{equation}
where 
\begin{equation}j_p= \int^{1/(2f_{\pi})}_{-1/(2f_{\pi})} \exp \{ i \pi \left[ f_{\pi } t ~\left( \bmod ~1 \right) \right] -2 \pi i p f_{\pi} t \}  f_{\pi} dt.
\label{eqn:jp}
\end{equation}
When $B_0$ is chosen such that $\Omega_0 = \pi  f_{\pi }$, the $p = 1$ term dominates, the $B_+$ field response is maximized, and $S_+$ becomes
\begin{equation}
S_+ = \frac{\Omega _{\perp}S_z\left|j_1\right|}{\Gamma } \exp \left( i \omega _{\pi} t + i \phi_{\pi} + i \alpha - i \frac{\pi}{2}  \right),
\label{eqn:splus}
\end{equation}
where $\alpha = \arg \left( j_1 \Omega_+\right)$, and $\Omega_{\perp} = \left|\Omega_x + i \Omega_y\right|$.
Calculating $j_1 = -2 i / \pi$, we can find the spin projection on the direction of the probe propagation \axis{x}:
\begin{equation}
S_x=
-\frac{2 \Omega_x S_z}{\pi \Gamma } \cos \left(\omega _{\pi }t + \phi_{\pi } \right) + 
\frac{2 \Omega_y S_z}{\pi \Gamma } \sin \left(\omega _{\pi}t + \phi_{\pi } \right).
\label{eqn:sx}
\end{equation}
The probe polarization rotation signal thus can be synchronously detected at $\omega_{\pi}$, and the components corresponding to $B_x$ and $B_y$ are orthogonal.
Note that the $B_y$ signal has a non-zero average, and the square wave demodulation shown in \fig{fig:setup}(b) discards the DC Fourier component of the signal. 
This suppresses the additive \oneOverF{} technical noise, but degrades the gain in $B_y$ channel by a factor of $\sim$2.4 compared to the $B_x$ channel.

Qualitatively, the shapes of the \axis{x} and \axis{y} signals [\fig{fig:setup}(b)] can be understood as follows.
Consider an ensemble of spins initially polarized along \axis{z} in zero net magnetic field. 
A small applied field $B_x$ or $B_y$ generates components of spin polarization along \axis{y} or \axis{x}, respectively, as described in~\cite{firstSERF}.  
Now superimpose an additional static field $B_z$; the spins begin to precess about \axis{z}.  
Adding an infinitely short \pipulse{} parallel to $B_z$ after $\pi$ radians of precession effectively eliminates half of each precession cycle.
Thus, a $B_x$ field generates static time-average polarization along \axis{y}, which sweeps out an arc about \axis{z} by $\pm \pi / 2$ radians, at the \pipulse{} repetition rate; likewise, a $B_y$ field generates a time-average static polarization along \axis{x} which sweeps out an arc about \axis{z} by $\pm \pi / 2$ radians.

In order to determine how small variations $\Delta B_0$ in the leading field affect the magnetometer performance, we find the modified $A'_{(+,p)}$ expression by substituting $\Omega_0 = \pi f_{\pi} + \delta$, $\delta = \gamma \Delta B_0,\, \delta \ll \pi f_{\pi}$ into Eq.~\ref{eqn:resonance}:
\begin{equation}
\begin{split}
A'_{(+,p)} = & \frac{-i \Omega _+S_zj_p}{\Gamma -i \left(\Omega _0+\pi  f_{\pi } + \delta -p \omega _{\pi }\right)} \simeq \\
& \simeq A_{(+,p)} \left(1 + \frac{i\delta}{\Gamma}\right) \simeq A_{(+,p)} e^{i \delta / \Gamma}.
\end{split}
\label{eqn:B0_deviation}
\end{equation} 
Similarly, small variations in the \pipulse{} area $A_{\pi} = \pi + \delta, \, \delta \ll \pi$ result in the modified $A'_{(+,p)}$ expression:
\begin{equation}
\begin{split}
A'_{(+,p)} = & \frac{-i \Omega _+S_zj_p}{\Gamma -i \left[\Omega _0 + (\pi + \delta)  f_{\pi } -p \omega _{\pi }\right]} \simeq \\
& \simeq A_{(+,p)} \left(1 + \frac{i\delta f_{\pi}}{\Gamma}\right) \simeq A_{(+,p)} e^{i \delta f_{\pi} / \Gamma}.
\end{split}
\label{eqn:A_deviation}
\end{equation} 

Eqs.~\ref{eqn:B0_deviation} and \ref{eqn:A_deviation} suggest that when the offset field $B_0$ deviates from the resonance condition (Eq.~\ref{eqn:resonance}), or the pulse area $A_{\pi}$ deviates from $\pi$, the magnetometer's sensitive axes rotate by the angle $\Delta \phi = \gamma \Delta B_0/\Gamma$ and $\Delta \phi = \gamma \Delta A_{\pi} f_{\pi} /\Gamma$, correspondingly.

\section{Experimental setup}

\begin{figure}[t]
  \centering
  \includegraphics[width=\columnwidth]{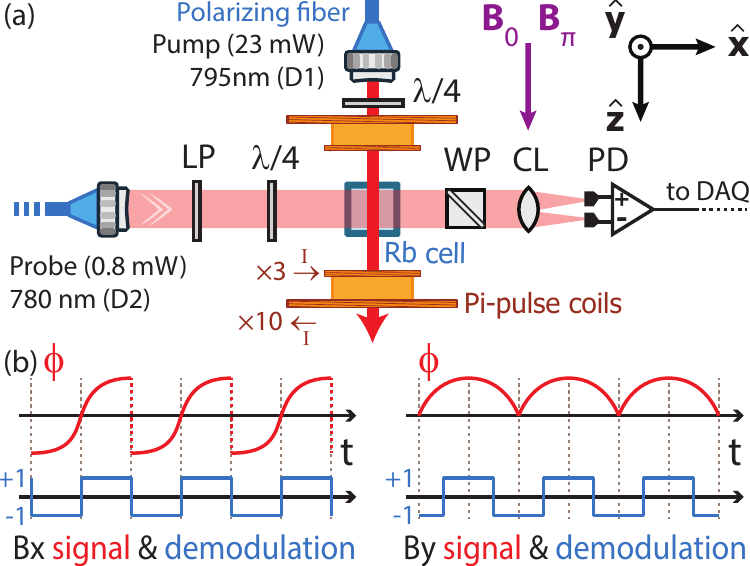}
  \caption{
  a) Experimental setup. LP -- linear polarizer, WP - Wollaston prism, CL -- condenser lens, PD --  differential polarimeter, DAQ -- data acquisition system. Not shown: heaters, field coils, and the magnetic shielding. 
  b) \pipulse{} magnetometer PD response to constant positive \axis{x} and \axis{y} fields, and the corresponding demodulation waveforms (illustration)}
  \label{fig:setup}
\end{figure}

The experimental setup [\fig{fig:setup}(a)] allows for direct comparison between the DC SERF, the Z-mode SERF and the \pipulse{} SERF magnetometers' performance.
We have independently optimized each magnetometer's parameters, and have found that all three have the largest optical gain at the same laser tuning and power.

The core of the setup is a rectangular vapor cell~(10$\times$10$\times$30~mm, \rubidium{}\,+\,165~Torr N$_2$) with two clear optical axes for the pump and probe beams, which propagate through the short dimensions of the cell. 
The cell is enclosed by a set of high-resistance ceramic heaters with counter-propagating wire traces in order to minimize stray magnetic fields, similar to~\cite{bulatowicz2012temperature}.
Heat-insulating padding made from 10~mm-thick aerogel sheets is placed between the heater assembly and a 3d-printed plastic housing.
The housing also provides frames for rectangular $B_x$ and $B_y$ field coils (37~mm~$\times$~32~mm, $\Delta$x=38~mm).
The AC field for the Z-mode SERF, $B_0$ and \pipulse{}s are created with a larger auxiliary coil system to improve the field uniformity~(see the Appendix).
The vapor cell is heated to approximately \temp{175} with AC at 401.5~kHz, chosen to minimize aliasing of the interference from the current in the heating elements into the demodulated signal~\cite{WyllieThesis}.

The linearly polarized probe beam (780~nm, 800\uW{}, \axis{x}) is delivered into a 4-layer $\mu$-metal shield via a polarizing single-mode fiber~(IXfiber $\lambda$=780~nm, $\diameter$125~$\mu$m core).
The optical frequency is tuned to the blue side of the D$_2$ line, and adjusted to maximize the magnetometer response.
After the fiber, the probe polarization is additionally cleaned up with an absorptive linear polarizer (LP), and the residual birefringence in the vapor cell walls is compensated with a $\lambda/4$ wave plate.
The probe polarization rotation is measured with a balanced differential polarimeter consisting of Wollaston prism (WP), a condenser lens (CL), and a matched photodiode pair (PD).
A differential current amplifier~\cite{hobbs} converts the PD difference current into voltage, which is then acquired and demodulated by the data acquisition system (DAQ) in real time.

The pump beam (23\mW, 795~nm, \axis{z}) is delivered into the magnetic shield via the same type of fiber as the probe beam, and is circularly polarized before entering the vapor cell.
The optical frequency is locked on the red side of the D$_1$ line, and fine-tuned to maximize the magnetometer response.
Light-shift gradients are minimized by operating the magnetometer in the diffusive SERF regime~\cite{Sulai2013} with high light intensity within the pump beam ($w_0$=0.3~cm) to ensure that the pumped atoms remain primarily polarized along \axis{z}.
Suppressing AC Stark shift gradients is particularly important in the \pipulse{} magnetometer setup, as they cause non-uniformity of $\boldvec{\Omega}_0$ across the cell volume. 
This broadens the magnetic resonance (Eq.~\ref{eqn:resonance}), reduces the response amplitude, and introduces  transients into the signal as the atoms precess out of phase with each other.
Atoms diffusing outside of the pump beam dominate the magnetic signal, unaffected by the light shifts and broadening.
With a relaxation rate $\Gamma$=435~s$^{-1}$ limited by Rb-Rb spin-destruction collisions, and a diffusion coefficient estimate of $D$=7~mm$^2$/s, the atoms traverse $\Lambda$=$2 \pi \sqrt{D/\Gamma}$=\,3~mm before being depolarized.

Pump laser power and frequency are stabilized with two PID controllers implemented in the magnetometer FPGA code, ensuring that the feedback is synchronous with the magnetic data acquisition.
The uncoated front surface of the magnetometer cell serves as a power pickoff, enabling monitoring of the pump power noise immediately before the cell.
The pump power is measured via a ceramic photodiode placed inside the magnetic shields, with the error signal fed back to a liquid crystal modulator (Meadowlark Optics D3060HV), stabilizing the pickoff light power.
The pump laser frequency is locked to a feature in the transmission signal of an auxiliary vacuum saturated absorption spectroscopy cell, which contains natural abundance rubidium.
The photocurrents of both the saturated absorption system and the pump power pickoff are amplified with SRS~570 current-to-voltage converters.

The polarimeter signal is digitized by a 16-bit ADC at 500~ksps synchronously with the \pipulse{} control signal produced by the FPGA~(NI-7851R).
The demodulation is performed in real time by multiplying the ADC data with a square wave [\fig{fig:setup}(b)].
The demodulated raw $B_x$ and $B_y$ data is streamed to the host computer at the rates of 1~ksps  (Z-mode) or 500~Hz (\pipulse{}) per channel, where it is converted into magnetic field units.

\section{Noise analysis}

We determine the sensitivity of the \pipulse{} magnetometer to $B_0$ variations by applying a low-frequency sinusoidal magnetic field along $B_z$, and measuring the response in $B_x$ and $B_y$ channels of the \pipulse{} magnetometer.
The corresponding crosstalk coefficients are $B_x/B_z = 6 \times 10^{-3}$ and $B_y/B_z = 27 \times 10^{-3}$.
The discrepancy between the responses is caused by non-orthogonality between $B_y$, $B_x$ and $B_0$, as the coils producing these fields are located on different frames.

The DC magnetic fields in the setup are generated by custom-made current supplies~\cite{WyllieThesis}. 
Based on the noise density measurements at 0.1~Hz, 1~Hz, and 30~Hz, we extrapolate the \oneOverF{} noise to lower frequencies.
We estimate the added noise due to $B_0$ current drifts at 0.01~Hz in $B_0$, $B_x$, and $B_y$ to be  292\fthz{},  1.8\fthz{}, and  7.9\fthz{}, correspondingly.
Similarly, the estimated \oneOverF{} noise in $B_x$ and $B_y$ supplies is 117\fthz{} at 0.01~Hz.
Although $B_x$ and $B_y$ drifts exceed the technical noise floor at frequencies below 10~Hz, it is still possible to achieve high magnetic field sensitivity by either reducing the dynamic range of an individual sensor, or by employing an array of sensors sharing the bias field compensation, along with a low-current gradient compensation field.

The \pipulse{}s are generated with a home-made half H-bridge circuit, described in the Appendix.
To the leading order, the pulse area is proportional to the square of pulse time, and to the coil power supply voltage. 
The fractional noise of the coil supply voltage measured $3 \times 10^{-9}$~Hz$^{-1/2}$ at 1 Hz.
The timing jitter $t_j \approx 88$~ps RMS (250~ps peak-to-peak, $f_0=40$~MHz) on the \pipulse{} duration $t_{\pi} = 4.675$~$\mu$s (nominal) generates a fractional noise $N_{t_\pi}$:
\begin{equation}
N_{t_\pi}=\frac{(t_{\pi}+t_j)^2}{ t_{\pi} \sqrt{f_{\pi}} } \approx \frac{ 2 t_j t_{\pi}}{ t_{\pi} \sqrt{f_{\pi}} } = 1.7 \times 10^{-6} \, \mathrm{Hz}^{-1/2}. 
\end{equation}
Over the time scale of the measurements ($\sim 100$~s), $N_{t_\pi}$ exceeds typical center frequency drift in a quartz oscillator ($\sim 10^{-8}$)~\cite{quartz_clock}.
The \pipulse{} area noise is thus dominated by the short-term phase noise of the FPGA clock.
Noting that the \pipulse{}s generate the same total precession as $B_0$, we estimate the equivalent $B_z$ noise density induced by the \pipulse{} duration instability as $N_{t_\pi} \times B_0 = 75$\fthz{}.
Based on the cross-talk coefficients, the induced $B_x$ and $B_y$ noise is 0.4\fthz{} and 2.0\fthz{}, correspondingly.


\section{Results}


\begin{figure}
  \centering
  \includegraphics[width=\columnwidth]{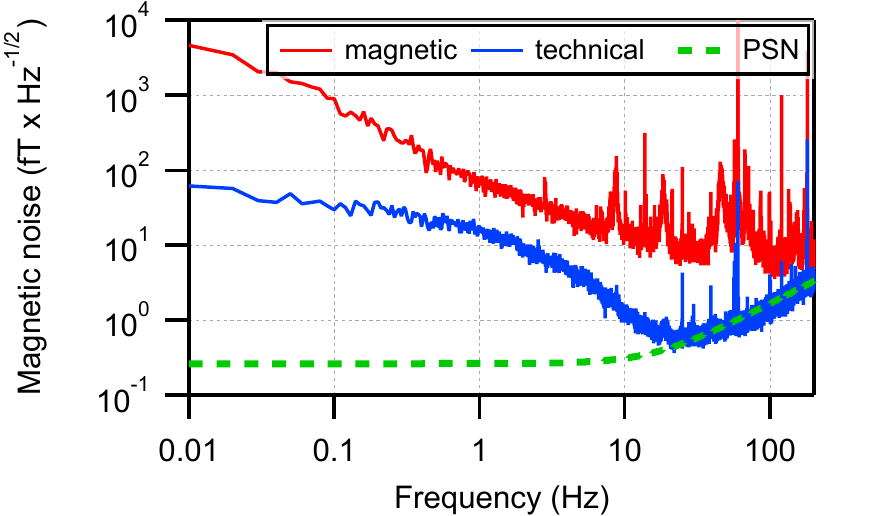}
  \caption{DC SERF magnetic (red), technical (blue), and calculated photon shot noise (green dashed line).}
  \label{fig:dc_serf}
\end{figure}

\begin{figure}
  \centering
  \includegraphics[width=\columnwidth]{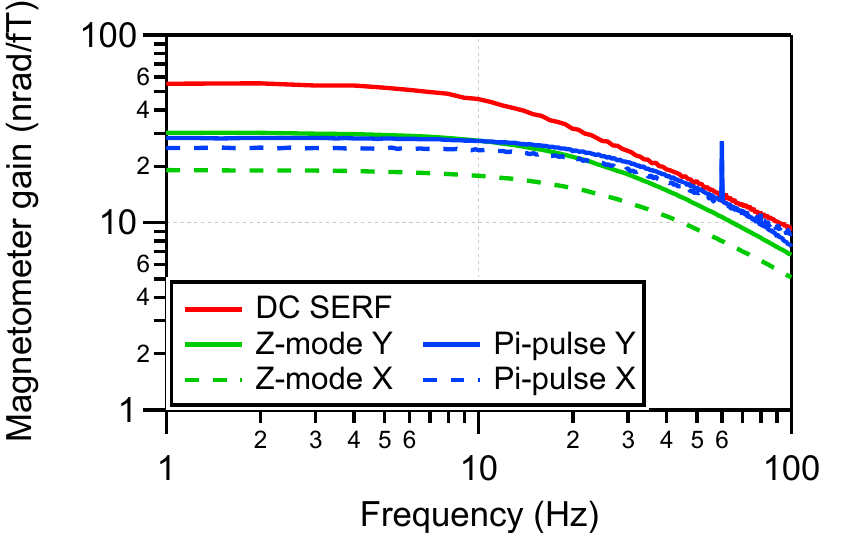}
  \caption{DC SERF (red), Z-mode SERF (green) and \pipulse{} (blue) magnetometer gains. }
  \label{fig:response}
\end{figure}

\begin{figure}
    \centering
    \begin{subfigure}{\columnwidth}
  	\includegraphics[width=\columnwidth]{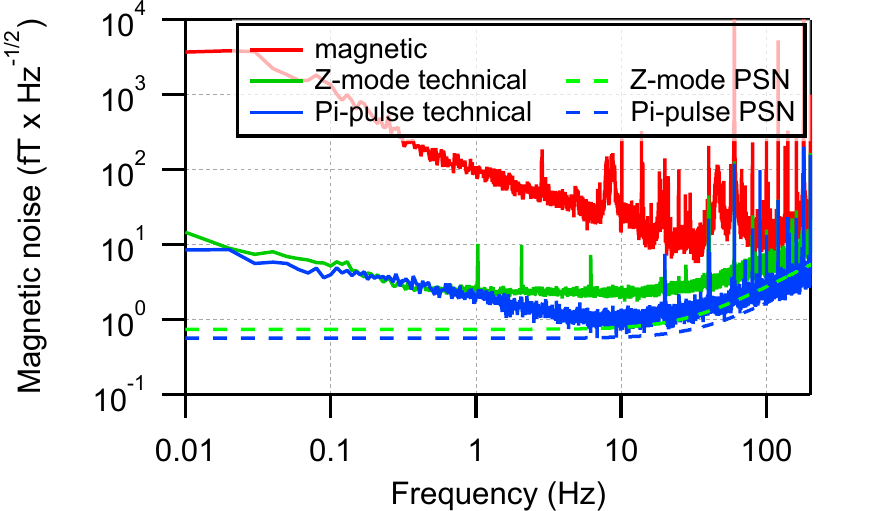}
  	\caption{\axis{x} axis}
  	\label{fig:noise_X}
    \end{subfigure}
    \\
    \begin{subfigure}{\columnwidth}
      \includegraphics[width=\columnwidth]{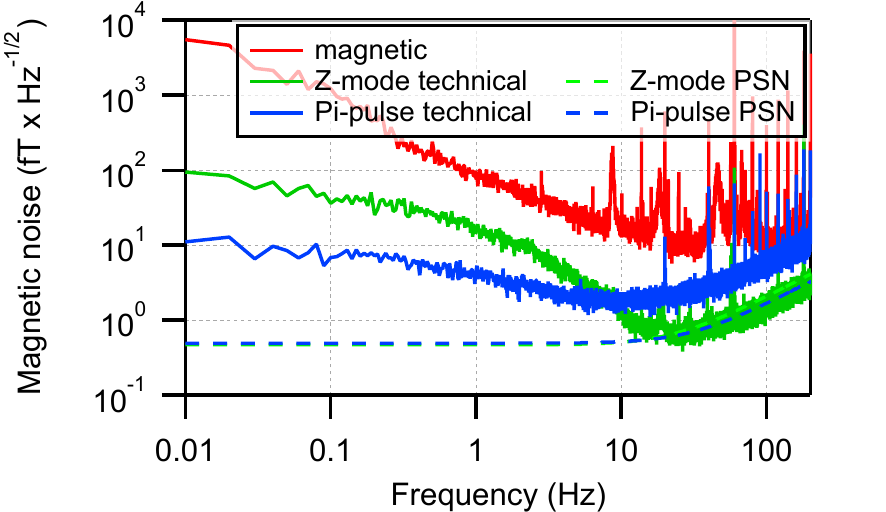}
  	\caption{\axis{y} axis}
  	\label{fig:noise_Y}
    \end{subfigure}
    \caption{Comparison of the \pipulse{} (blue) and Z-mode (green) magnetometers. The dashed lines represent corresponding photon shot noise limits. The magnetic noise (red) is measured with the Z-mode magnetometer.}
    \label{fig:zmod_vs_pi_pulse}
\end{figure}

We begin by implementing a DC SERF magnetometer in order to provide a performance baseline for our experimental setup.
The noise spectral density of a DC SERF optimized for the best technical noise performance is presented on \fig{fig:dc_serf}.
Each noise trace is created by averaging the spectra of several 100~s-long samples. 
The technical noise (blue) is calculated by adding the noise contributions from the probe, pump power, and pump frequency fluctuations in quadrature.
The photon shot noise (green dash) is calculated theoretically, and the magnetic noise (red) is the measured magnetic field noise in the setup.
The lowest magnetic noise measured in this setup is 10\fthz{}, limited by the Johnson noise of the magnetic shield.
The technical noise limit of the setup approaches the photon shot noise limit at frequencies above 25~Hz, attaining the minimum of 0.7\fthz{}.
Although the photon shot noise further decreases below this frequency, the magnetometer sensitivity still degrades due to the technical noise increase.

\begin{figure}
  \centering
  \includegraphics[width=\columnwidth]{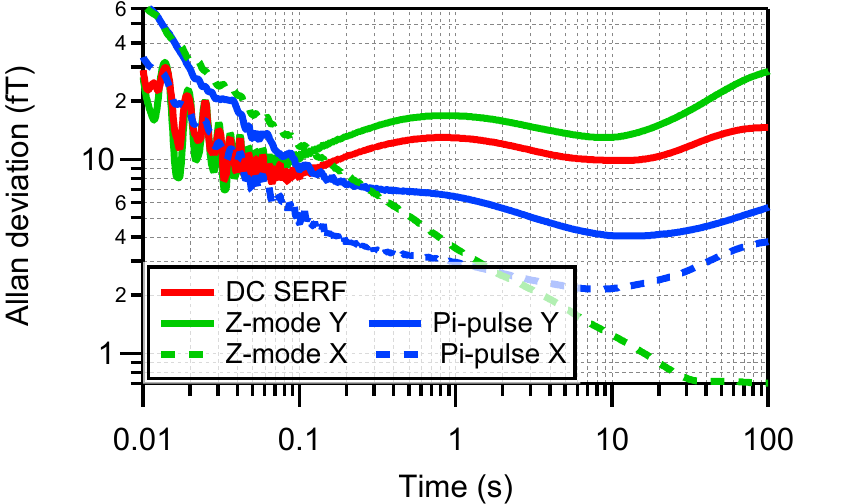}
  \caption{Allan deviation of DC SERF (red), Z-mode (green) and \pipulse{} (blue) magnetometers}
  \label{fig:adev}
\end{figure}

The optical gains of the DC SERF, Z-mode SERF and the \pipulse{} magnetometers are presented on \fig{fig:response}.
The gain is measured by applying a known magnetic field and measuring the optical rotation signal as a function of frequency~\cite{DeLandThesis}.
In the \pipulse{} magnetometer, $B_0$=\bmainValuePiPulse{} and the \pipulse{}s jointly generate full precession cycles at $f_{\pi}$=500~Hz. 
The gyromagnetic ratio is therefore $\gamma$=\gratioPiPulse{}, which corresponds to a polarization of p=\polarizationPiPulse{} in the spin-temperature limit~\cite{Happer1973}.
In the Z-mode SERF, the modulation amplitude (25~nT) is selected to maximize the $B_x$ sensitivity, while the modulation frequency matches the \pipulse{} repetition rate.

The technical noise comparison for Z-mode and \pipulse{}-mode magnetometers is presented on \fig{fig:zmod_vs_pi_pulse}.
In contrast to the DC SERF, Z-mode and \pipulse{} magnetometers have an improved low-frequency noise performance, except for the \axis{y} direction in Z-mode [\fig{fig:zmod_vs_pi_pulse}(b)], which does not benefit from the added modulation.
In addition, we assess the stability of each magnetometer using the Allan deviation (\fig{fig:adev}), computed from a set of 400~s-long data samples.
This provides an estimate of the magnetometers' performance when they are operated as a sensor array in a closed feedback loop.
The \pipulse{} magnetometer readout can be averaged for up to 10~s to attain a technical noise floor of 2.2~fT (\axis{x}) and 4.0~fT (\axis{y}), which is within the stability requirements of operating an fMCG array.
Z-mode \axis{y} and the DC SERF signals exhibit drifts at the time scales above 0.1~s, 
while the Z-mode \axis{x} signal has the lowest drift and is dominated by the sensor noise up to at least 40~s of integration time.


\begin{figure}
  \centering
  \includegraphics[width=\columnwidth]{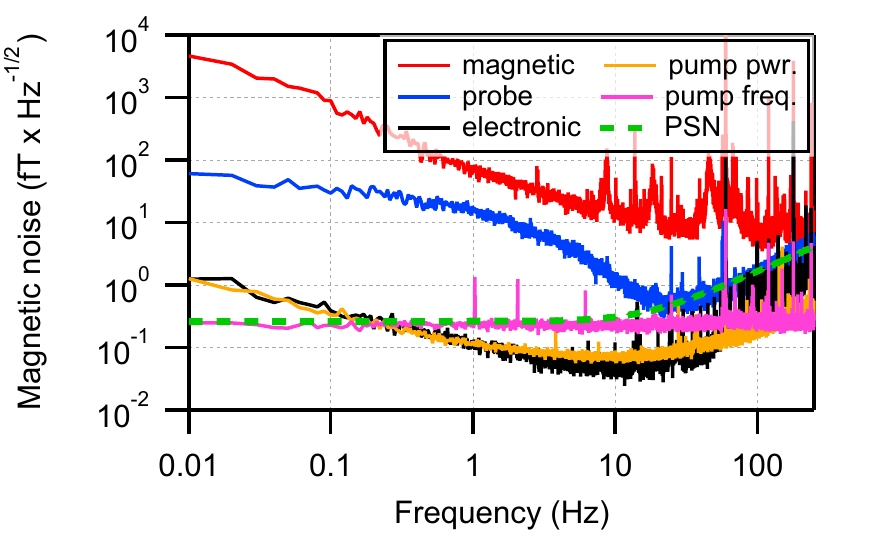}
  \caption{
  DC SERF magnetometer noise density
  }
  \label{fig:DC_SERF_noise}
\end{figure} 

\begin{figure}
    \centering
    \begin{subfigure}{\columnwidth}
      \includegraphics[width=\columnwidth]{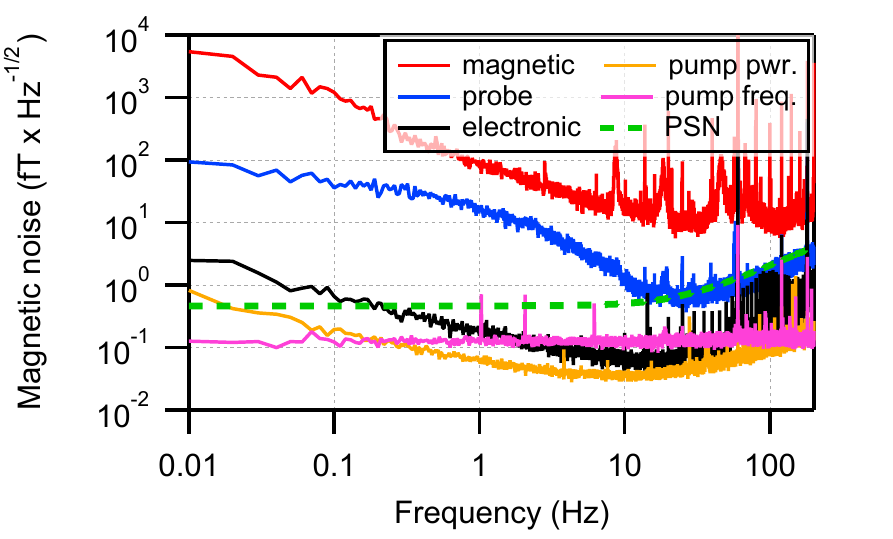}
      \caption{
      Y axis
      }
      \label{fig:Z_mode_Y_noise}
    \end{subfigure}
    \\
    \begin{subfigure}{\columnwidth}
      \includegraphics[width=\columnwidth]{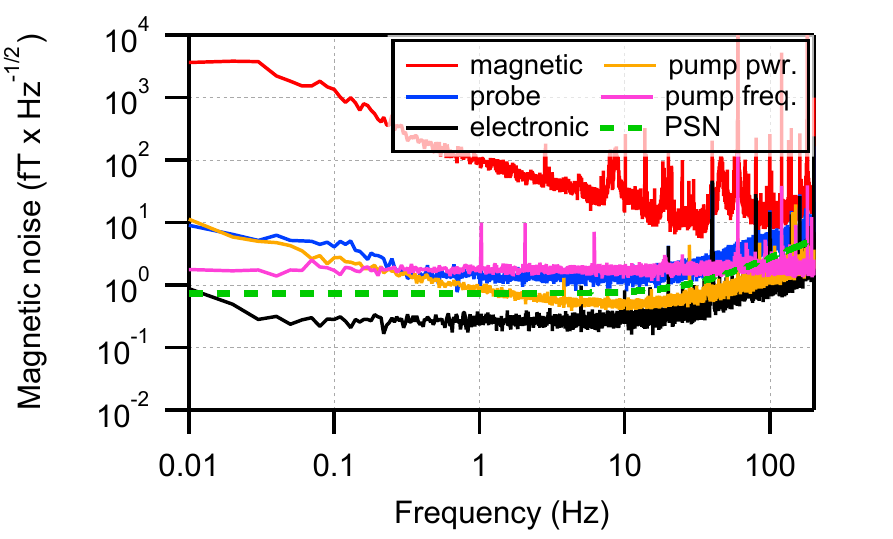}
      \caption{
      X axis
      }
      \label{fig:Z_mode_X_noise}
    \end{subfigure}
    \caption{Z-mode magnetometer noise density}
    \label{fig:Z_mode_noise}
\end{figure}

\begin{figure}
    \centering
    \begin{subfigure}{\columnwidth}
      \includegraphics[width=\columnwidth]{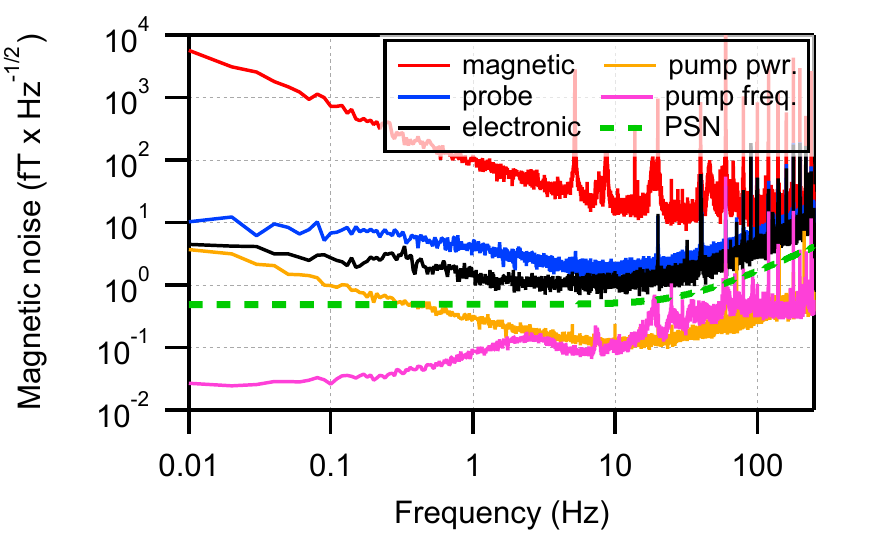}
      \caption{
      Y axis
      }
      \label{fig:Pi_pulse_Y_noise}
    \end{subfigure}
    \\
    \begin{subfigure}{\columnwidth}
      \includegraphics[width=\columnwidth]{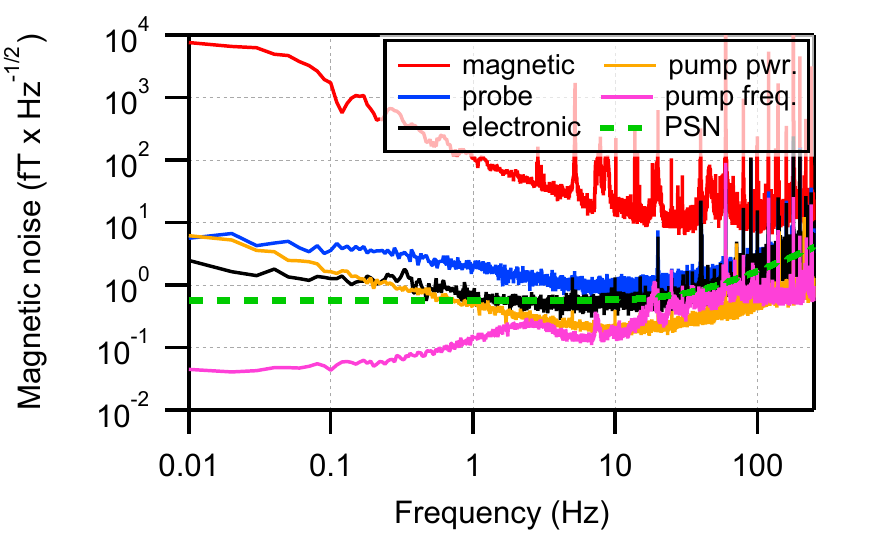}
      \caption{
      X axis
      }
      \label{fig:Pi_pulse_X_noise}
    \end{subfigure}
    \caption{Pi-pulse magnetometer noise density}
    \label{fig:Pi_pulse_noise}
\end{figure}

The detailed technical noise composition of each magnetometer is presented on \fig{fig:DC_SERF_noise} (DC SERF), \fig{fig:Z_mode_noise} (Z-mode SERF), and \fig{fig:Pi_pulse_noise} (\pipulse{}).
On these plots, the magnetic noise trace (red) shows fluctuations of the real magnetic field during the measurements.
The probe noise trace (blue) is the magnetic sensitivity limit imposed by the optical detection scheme.
It is measured by recording the magnetometer signal while blocking the pump laser beam.
The probe photon shot-noise (dash green) is calculated as $\rho = \sqrt{4 e I_{\mathrm{pd}}}$, where $e$ is the electron charge and $I_{\mathrm{pd}}$ is the current through a single photodiode of the balanced polarimeter. 
The electronic noise trace (black) is the sensitivity limit due to the electronic noise in the front-end amplifier, combined with the data acquisition and demodulation process.
The electronic noise is measured by recording the magnetometer signal with both pump and probe laser beams blocked.
The pump power (yellow) and frequency (pink) noise traces are calibrated by sequentially applying a sinusoidal ($f = 23$~Hz) modulation to each corresponding PID set point.
The power and frequency monitor readouts are captured simultaneously with the magnetometer signals.
By comparing the peak amplitudes at the calibration frequency in the magnetic signal and the readouts, we calibrate the readout signals into the magnetic field units.


\begin{figure}
    \centering
    \begin{subfigure}{\columnwidth}
      \includegraphics[width=\columnwidth]{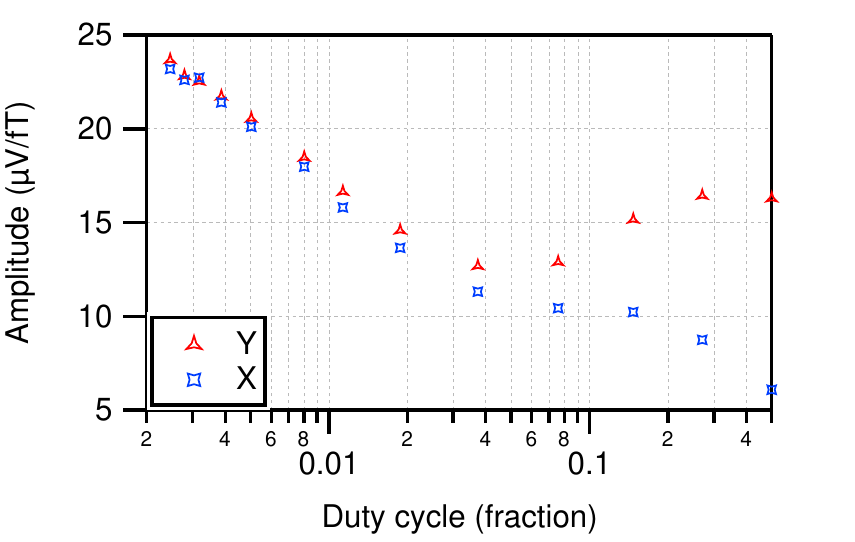}
      \caption{
      Magnetic response amplitude
      }
      \label{fig:A0_vs_duty_cycle}
    \end{subfigure}
    \\
    \begin{subfigure}{\columnwidth}
      \includegraphics[width=\columnwidth]{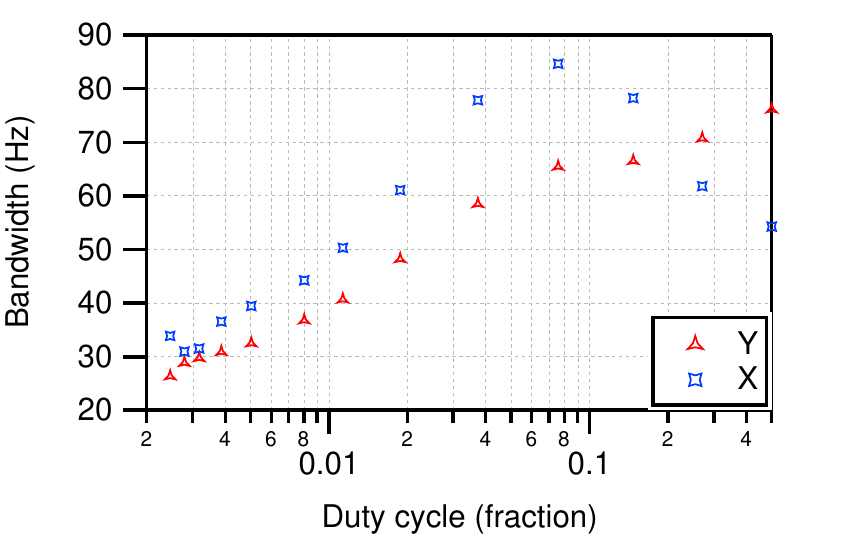}
      \caption{
      Magnetic response bandwidth
      }
      \label{fig:w0_vs_duty_cycle}
    \end{subfigure}
    \caption{Effects of the \pipulse{} duty cycle on the magnetometer's response}
    \label{fig:response_vs_duty_cycle}
\end{figure}

Although the peak \pipulse{} field amplitude ($16$~$\mu$T) is multiple orders of magnitude larger than the intended device sensitivity, it is still reachable, since the magnetic field only determines the instantaneous precession frequency of the atoms, while the magnetometer measures their phase.
The \pipulse{} magnetometer can operate in the SERF regime provided that the \pipulse{} duration is short compared to the time between spin-exchange collisions and the Larmor precession rate in $B_0$ is less than the spin-exchange-collision rate ($\sim 0.3 \times 10^6$~1/s).

The effect of spin-exchange collisions occurring during the \pipulse{} manifests as an additional spin-relaxation mechanism, resulting in decreased magnetic response magnitude and increased bandwidth.
In \fig{fig:response_vs_duty_cycle} we measure the magnetometer response while varying the \pipulse{} duty cycle between 0.25\% and 50\%.
During the measurements, both pulse duration and amplitude are adjusted to maintain the pulse-induced precession at $\pi$, while keeping $B_0$ constant.
Although the \pipulse{} duration increase adversely affects the magnetometer response, it only degrades by a factor of two as the duty cycle increases by an order of magnitude.
Even in the extreme case of 50\% duty cycle, the $B_x$ and $B_y$ responses are still measurable.
In this experiment the minimum duration of the \pipulse{} was limited by the largest \pipulse{} coil voltage that could be applied without damaging the electronics.

\section{Conclusion}
In this paper, we have demonstrated a new type of a spin-exchange relaxation-free magnetometer and  measured its technical noise. 
While overall it has lower gain compared to the traditional DC SERF magnetometer, it offers an important advantage of enabling synchronous detection in two directions simultaneously.
The \axis{x} and \axis{y} magnetic field signals are generated at the chosen \pipulse{} frequency and are orthogonal, minimizing the technical \oneOverF{} noise contribution from the laser and temperature drifts in both signals simultaneously. 
This is especially important in biomagnetic applications, since a significant fraction of the signal power is contained in 0.1--100\,Hz frequency range, where the measurement sensitivity is often limited by the \oneOverF{} noise in the detection system~\cite{WyllieThesis}. 
This also opens a possibility of precise gradiometry measurements with several independent sensors, as the improved long-term stability can be used for field stabilization and better cancellation of the environmental noise.
Our next goal is to implement a \pipulse{} magnetometer array for fMCG, and perform a direct low-frequency low-noise magnetic field gradient measurement.

\begin{acknowledgments}
This work was supported by the grant R01HD057965-05 awarded by the National Institutes of Health, as well as the grant PHY-1607439 by the National Science Foundation.
\end{acknowledgments}

\appendix*

\section*{Appendix}

\begin{figure}
  \centering
  \includegraphics[width=\columnwidth]{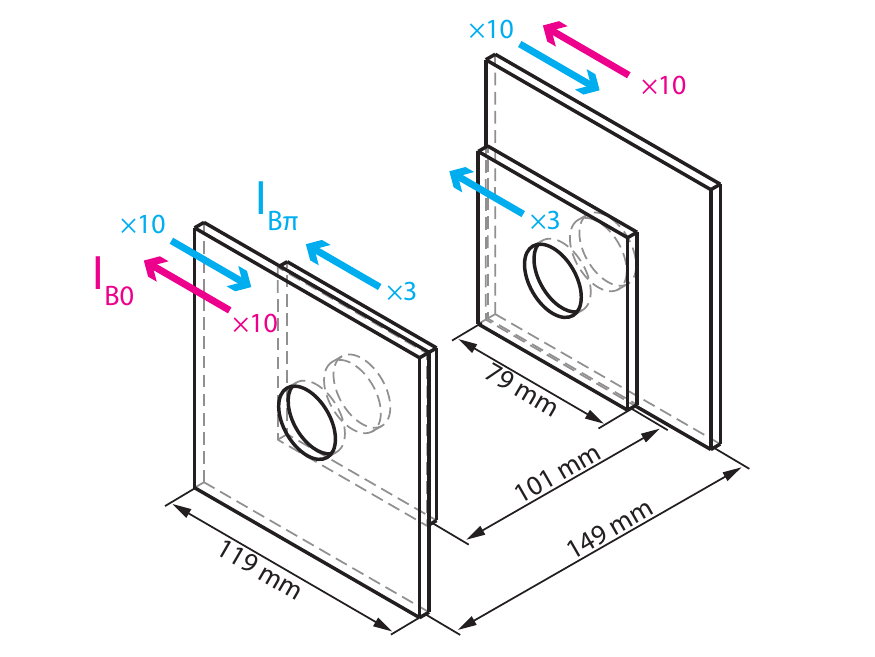}
  \caption{
   Auxiliary coil frame housing the \pipulse{} and the offset field coils
  }
  \label{fig:coils_schematic}
\end{figure}

\begin{figure}
  \centering
  \includegraphics[width=\columnwidth]{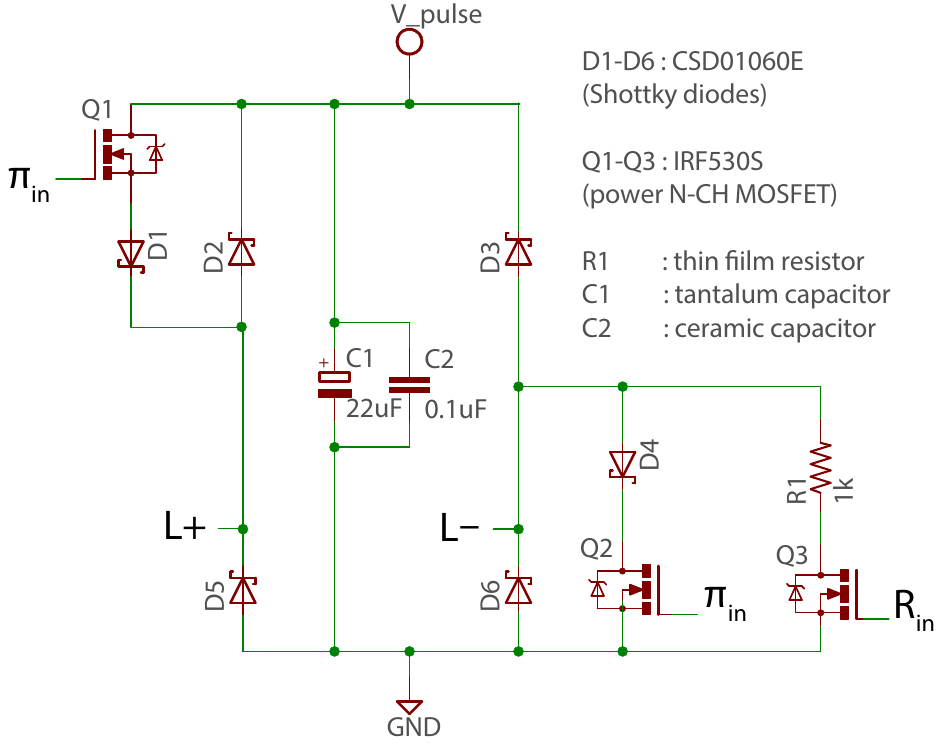}
  \caption{
   Schematic of the pulse circuit\footnote{The actual pulse circuit is a full H-bridge; in this experiment only one half was used.}. The MOSFET drivers~(MIC4420) and the corresponding power rail are not shown. $\pi_{\mathrm{in}}$ -- main pulse control signal; $\mathrm{R}_{\mathrm{in}}$ -- ringing suppression gate control signal; $\mathrm{L}+$ and $\mathrm{L}-$ -- coil leads; V\_pulse -- pulse power rail; GND -- ground.
  }
  \label{fig:pulse_circuit}
\end{figure}

%

The offset field and the \pipulse{} coils are wrapped on top of each other on the auxiliary coil frame~(\fig{fig:coils_schematic}), and consist of two square coils~($L = 119$~mm), $N = 10$ wraps each, separated by $\Delta x = 149$~mm.
The \pipulse{} coils have an additional compensation coil set~($L = 79$~mm, $\Delta x = 101$~mm, $N=3$), wrapped in the opposite direction to the primary coils.
The \pipulse{} coil geometry is designed to minimize the $dB_z/dz$ field gradients along the pump axis.
Calculated current-to-field conversion coefficients are $\beta = 31$~nT/mA for the \pipulse{}, and $\beta = 55$~nT/mA for the offset field coil.

The \pipulse{} control signals are generated by NI-7851R FPGA, and converted to current pulses via a custom half-H-bridge circuit (\fig{fig:pulse_circuit}).
The power rail V\_pulse controls the pulse amplitude; it is connected to a low-noise HP6205C DC power supply.
During the \pipulse{}, transistors Q1 and Q2 are switched ``on'', connecting V\_pulse and GND to the coil leads. 
The current through the \pipulse{} coils increases approximately linearly with a slope determined by V\_pulse.  
Capacitors C1 and C2 provide an additional low-impedance source for the pulse current and a sink for the return current, helping to minimize the dynamic loading of the power supply.
Schottky diodes with zero reverse recovery time and minimal parasitic capacitance are chosen for this application, in order to reduce undesirable current oscillations through the \pipulse{} coil. 
Further suppression of the current oscillation is achieved by activating a supplementary ringing suppression gate Q3 shortly after completion of the main pulse, when the current through the \pipulse{} coil reaches zero. 
This shunts any residual energy stored in the parasitic capacitances of the circuit and the magnetic field to ground through resistor R1.
This prevents the oscillations of the residual energy between the \pipulse{} coil and the parasitic capacitances that would otherwise have occurred. 
The value of resistor R1=$\sqrt{L/C_{\mathrm{Q3}}}$ is chosen to balance energy dissipation rate against oscillations of the current during the \pipulse{}.  
Optimal timing and duration of the Q3 control signal are determined by connecting high impedance scope probes to the coil leads and verifying that the ringing after the pulse is minimized.
The \pipulse{} control signals are buffered with MOSFET drivers (MIC4420, not shown), which are powered with HPE3620A DC power supply.  
Optionally, the control signals may be buffered with a bridge driver to increase the maximum allowed V\_pulse.


To assess the magnetic field pulse shape, we connect a 1~$\Omega$ resistor in series with the \pipulse{} coil and measure the voltage drop across the resistor throughout the pulse.
With V\_pulse=14~V, the current linearly rises from 0~A to 0.5~A over 4.7\uS{} during the pulse active phase, and drops back to 0~A over 2.7\uS{} after the pulse completion.
This provides an estimate for the coil inductance $L=130$~$\mu$H, and the peak magnetic field $B = 16$~$\mu$T.
With $\gamma=5.6$~Hz/nT, we can calculate the pulse area $A=2.1$~rad, which is within the order of magnitude of $\pi$.

\bibliography{bibliography.bib}

\end{document}